\documentclass{llncs}
\usepackage{graphicx} 

\usepackage[english]{babel}
\usepackage[utf8x]{inputenc}
\usepackage[T1]{fontenc}

\usepackage{xcolor,comment,afterpage}
\usepackage[hyphens]{url}
\usepackage{csquotes}

\newcommand{\antero}[1]  {\noindent \textcolor{cyan}{[#1 -Antero]}}
\newcommand{\tommi}[1]   {\noindent \textcolor{blue}{[#1 -Tommi]}}
\newcommand{\cesare}[1]  {\noindent \textcolor{green}{[#1 -Cesare]}}

\title{On the Future of Software Reuse\\in the Era of AI Native Software Engineering}
\author{Antero Taivalsaari\inst{1}, Tommi Mikkonen\inst{2} \and Cesare Pautasso\inst{3}}

\institute{
Nokia Technologies, Tampere, Finland\\
\email{antero.taivalsaari@nokia.com} \and
University of Jyväskylä, Jyväskylä, Finland\\
\email{tommi.j.mikkonen@jyu.fi} \and
Software Institute, Università della Svizzera italiana (USI), Lugano, Switzerland\\
\email{c.pautasso@ieee.org}}

\begin{document}

\maketitle

\begin{abstract}
\noindent
  Software development is currently under a paradigm shift in which artificial intelligence and generative software reuse are taking the center stage in software creation. Earlier opportunistic software reuse practices and organic software development methods are rapidly being replaced by "AI Native" approaches in which developers place their trust on code that has been generated by artificial intelligence.  This is leading to a new form of software reuse that is conceptually not all that different from cargo cult development.  
  In this paper we discuss the implications of AI-assisted generative software reuse, bring forth relevant questions, and define a research agenda for tackling the central issues associated with this emerging approach.\\
  \\
  \textbf{Keywords}: Software engineering, software reuse, AI native software engineering, generative software reuse, software 2.0.
\end{abstract}

\section{Introduction}


In the past twenty-five years, the way people develop software has been affected profoundly by the World Wide Web.  The emergence of the Software as a Service (SaaS) model \cite{turner2003turning,artofsaas20115}, Internet-based developer forums, and open source software repositories such as GitHub, Node Package Manager (NPM) and Python Package Index (PyPI) -- with literally millions of available open source software components and libraries -- have enabled an approach in which software developers routinely trawl the Internet for ready-made solutions online.  Component selections are commonly performed based on popularity ratings alone, and the discovered libraries and code snippets are included in applications with little consideration or knowledge about actual technical details or quality. This approach is \emph{all about combining unrelated, often previously unknown software artifacts by joining them with "duct tape and glue code"} \cite{hartmann2008hacking}. Depending on the viewpoint and desired connotation, such development is referred to as \emph{opportunistic design} \cite{hartmann2008hacking}, \emph{opportunistic reuse}, \emph{ad hoc reuse}, \emph{scavenging} \cite{krueger1992software}, \emph{software mashups}, \emph{mashware} \cite{mikkonen2010mashware}, or sometimes even "\emph{frankensteining}" \cite{hartmann2008hacking}.  The resulting approach bears the imprint of \emph{cargo cult development} or \emph{cargo cult programming} \cite{Lip04} -- the ritual inclusion of program structures for reasons that the programmers do not fully understand. We have discussed opportunistic design and its implications for software development in a number of earlier papers \cite{taivalsaari2019programming,OpportunisticDesignTaivalsaariIEEE2019,makitalo2020opportunistic}. 

In recent few years, Artificial Intelligence and Machine Learning (AI/ML) have taken the center stage in the evolution of software reuse.  While Artificial Intelligence is actually a rather old research topic, originating from pioneering work by Alan Turing, Marvin Minsky, Dean Edmonds, John McCarthy and many others in the early 1950s, the real renaissance and boom in AI/ML technologies did not start until the 2010s.  The rate at which AI/ML technologies have evolved in the past fifteen years has far surpassed even the wildest expectations and predictions that could have been made by the forefathers of these technologies.  Just like the emergence and the subsequent popularization of the Internet in the turn of the millenium, the resurrection of AI/ML technologies is now reshaping various industries and has already resulted in a profound impact on many walks of life, including software development.



In general, software reuse and software development more broadly are currently under a paradigm shift in which Artificial Intelligence (AI) -- in particular Generative AI \cite{ebert2023generative} -- has taken an increasingly central role in assisting developers in their software creation activities.  This is effectively leading to a new form of software reuse in which automatically generated "AI native" software artifacts form the basis for software systems. In the new generative reuse approach, developers are requesting AI-based assistants to generate code for them, unlike in the past when developers were manually searching for pre-existing software components from libraries and code repositories such as the aforementioned GitHub, NPM and PyPI systems. These AI-generated artifacts can range from small code snippets and module fragments to comprehensive application skeletons or in some cases fully functional applications or even complete end-to-end systems.  This new model of software development is sometimes referred to also as \emph{software 2.0}\footnote{\url{https://karpathy.medium.com/software-2-0-a64152b37c35}}.

This new generative approach to software reuse has already resulted in a considerable mental model change for developers.  Basically, developers must be capable enough to judge the quality and suitability of the artificially generated code to the task at hand.  This can in many cases be very challenging, since AI assistants may use innumerable codebases as training datasets to derive the automatically generated designs.  Furthermore, the generated code may no longer reflect the mental model of any human developer; rather, it can be an amalgamation of various different development styles that have been conceived from several unrelated but seemingly similar solutions.  Generated code can in many cases be nearly perfect for the intended purpose.  However, it is not uncommon for AI to hallucinate and generate results that will look deceivingly convincing to an untrained eye but that are actually erroneous or even completely bogus. 

In many ways, generative reuse can be viewed as a new form of \emph{cargo cult development} -- ritual inclusion of code and program structures that originate from external sources without consideration or adequate understanding of the purpose, relevance or the potential side-effects of the borrowed artifacts \cite{feynman1998cargo,Lip04,maki2019cargo,taivalsaari2019programming}. Just like in cargo cult development, developers are (re)using code that they do not necessarily understand at all.  In the classic cargo cult programming scheme, developers are blindly doing something simply because others have used a certain piece of code or certain development approach earlier -- basically placing their trust on artifacts that have already been known to work in other contexts.  In contrast, in AI-based generative reuse developers place their trust on code that is generated by an external "oracle" whose inner workings are in essence completely unknown to the user.

In this paper, we discuss the implications of this new model of generative reuse and AI native software engineering based on our cumulative experience of over hundred years in the software engineering domain.  We start the paper with a brief history of software reuse in the past sixty years in Section 2, followed by an overview of the emerging era of AI native software engineering in Section 3.  We will then discuss the implications of AI assisted development on software reuse, followed by some discussion, tentative conclusions and predictions on the near term evolution as well as possible longer-term impacts on software development in Section 4.  In Section 5, we will summarize relevant research topics and questions that can serve as a research agenda and call for action for tackling the central issues associated with this emerging approach.  Finally, Section 6 concludes the paper.

\section{Brief History of Software Reuse}

\vspace*{1mm}
\begin{quote}
\emph{“The biggest difference between time and space\\
is that you cannot reuse time”}\\
\hspace*{4mm}-- Merrick Furst
\end{quote}

By definition, \emph{reuse is the act of using something that has already been used earlier}.  In essence, reuse refers to \emph{repeated use}: using the same physical or virtual artifacts multiple times -- minimally at least twice but commonly over and over again and possibly in many different contexts.

General purpose, commercially available software component libraries have been proposed ever since the famous NATO 1968 conference in which the term \emph{software engineering} was also introduced \cite{nato1968}. Doug McIlroy's pioneering 1968 paper that was included in the proceedings of the aforementioned conference called for \emph{high-quality mass produced software components to be used in large industrial scale} \cite{McI68}.  As a research topic, software reuse became especially popular in the 1980s \cite{Jon84,LaG84,LSW87,BiR87}, following the successful workshop on software reuse arranged by ITT Corporation in September 1983 \cite{itt1983proceedings}. An extensive survey on the state of the art in software reuse in the mid-1980s is provided by Jones \cite{Jon84}.  In practice, commercial success of larger-scale software reuse and component libraries did not begin until the 1990s, though. 

Although the \emph{potential} for software reuse was high in the 1980s and early 1990s, actual reuse rates remained very low.  Those days developers actually preferred writing their own code, and took pride in doing as much as possible from scratch. In fact, they were effectively expected or forced to do so, since third-party components were not widely available or easy to find. Before the advent of the World Wide Web, trade magazines, journals and word-of-mouth were the primary sources for advertisements, recommendations and reliable reviews or feedback on third-party components.  

Furthermore, before the widespread adoption of open source software development, components were rarely available for free or with such license terms that would favor large-scale use above and beyond those libraries that were provided as part of operating systems and their built-in programming environments.  In addition, there were technical hindrances as well.  Before the popularization of object-oriented programming languages and methods, software systems simply did not have good enough interfaces that would have made software reuse feasible in a larger scale, in spite of the pioneering work that had been carried out on software modularity by Parnas, Liskov, Snyder, Zilles and others earlier in the 1970s \cite{Par72a,Par72b,Zil73,Par79,LSA77}.

It was not until the mid-1980s and 1990s that software reuse started becoming more realistic in a larger scale.  Companies such as Borland, Metrowerks and Microsoft started expanding their C, C++ and Pascal programming language compiler offerings by bundling increasingly powerful libraries into their Integrated Development Environments (IDEs). Prior to that, most language compilers arrived with only a minimal set of libraries such as POSIX libraries for the C language on Unix, and many such libraries tended to have dependencies that limited their use only to specific computing platforms or specific hardware.  The Java programming language -- originally introduced in May 1995 -- was the first widely adopted system to include a comprehensive set of platform-independent APIs for nearly all basic purposes as a built-in feature of the programming language itself, enabling software reuse in a much larger scale across different computing platforms and different types of computing devices. However, even with the Java platform, its original catchy "\emph{write once, run everywhere}" premise ultimately failed as the Java platform splintered into different factions targeting different types of computing environments ranging from smartcards and mobile devices to desktop workstations and enterprise servers.

\begin{figure}[h]
    \centering
    \includegraphics[width=0.99\linewidth]{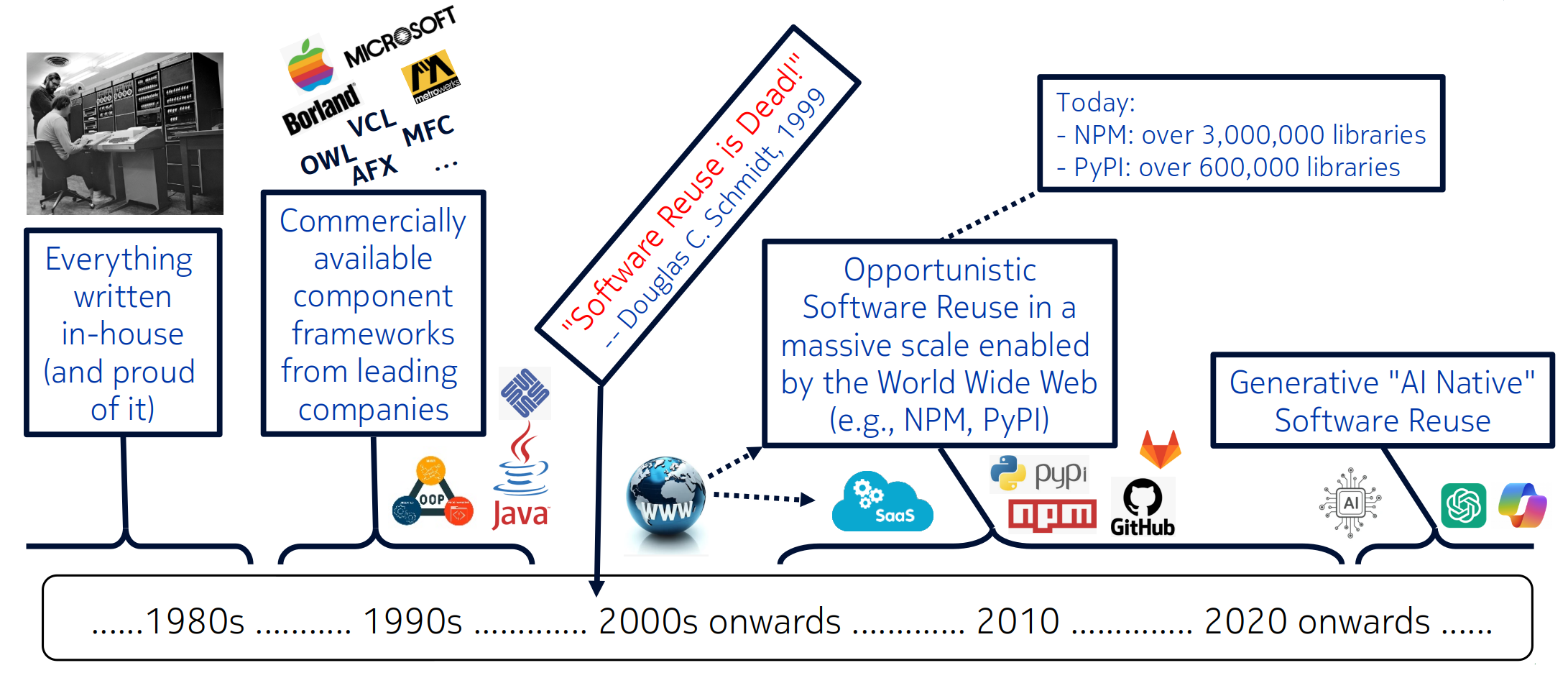}
    \caption{Simplified History of Software Reuse}
    \label{fig:history}
\end{figure}

Figure \ref{fig:history} presents a simplified timeline of software reuse depicting the evolutionary steps summarized above.  An interesting anecdote is that at one point \emph{just before the arrival of the opportunistic reuse enabled by the World Wide Web, some researchers had already proclaimed software reuse to be dead} \cite{schmidt1999software}.

Approximately from year 2000 onwards, the World Wide Web started changing things dramatically.  Although the World Wide Web had its origins in the early 1990s as a relatively simple document sharing environment, the Web gradually evolved into a dominant mass distribution platform for all things digital -- including also software components and libraries.  Software component repositories such as Node Package Manager and Python Package Index grew rapidly in the 2010s from providing thousands of components to millions of components.  At the time of this writing, the Python Package Index offers over 600,000 libraries, and NPM over three million packages (!) -- most of them with a permissive open source MIT license\footnote{\url{https://opensource.org/license/mit}}. 

Coincidentally, while all the programming features on the World Wide Web -- such as the introduction of the JavaScript programming language in the mid-1990s -- were largely an afterthought \cite{DeathOfBinary,ObjectsInTheCloud}, the evolution of the World Wide Web also led to the \emph{Software as a Service} (SaaS) model \cite{turner2003turning,artofsaas20115} in which the earlier "shrink wrapped" native applications were replaced by software that was provisioned (and in many cases run) directly from the World Wide Web.  Such web-based "evergreen" (always up-to-date) applications do not require any manual installation, and they can be updated worldwide in seconds, giving them a massive competitive edge over traditional binary applications that require explicit installation and updates by the users themselves.


Soon after the turn of the millennium, a tipping point was reached. Nowadays, it is nearly impossible to write any significant software systems without reusing third-party components extensively.  In essence, the developers themselves commonly write only the "tip of the iceberg" while the bulk of the system comes from external sources and unknown developers -- and typically with a lot of dependencies with additional subcomponents written by other unknown developers.  Today's software reuse scene is effectively all about hordes of software developers producing and reusing a cornucopia of open source components of varying and often unknown quality. While such an approach can be very convenient for developers, such form of reuse is very \emph{ad hoc} in its practices compared to the systematic textbook methodologies that were originally proposed for reuse two or three decades ago \cite{LaG84,kim1998software}.  More broadly, this is dramatically different from software development in the 1980s and 1990s when the developers had little choice except to write nearly all of the code themselves.



To give one concrete example, one area that has been profoundly impacted by the emergence of open source software is \emph{backend development of cloud-based systems}.  Thirty years ago -- at the beginning of the Internet boom in the mid-to-late 1990s -- if one were to establish a cloud-based Internet service, each company had to create their own custom solution.  Typically, each company bought the biggest server(s) that they could afford (for instance, a Sun Microsystems E10000 server or perhaps even a few of them for redundancy), and then installed web server software and the Java Enterprise Edition (J2EE) development stack on these machines. Nearly all the other software had to be written from scratch. 

Today, in contrast, open source components for backend development abound, and cloud system developers rarely have to write any of the major components themselves. Consequently, backend system development is mainly about picking the most applicable open source components and configuring  those components to be hosted and orchestrated in public or private cloud infrastructures. This becomes even simpler with \emph{serverless computing} in which most of the component selection and configuration work is abstracted away \cite{Hassan2021}.

\section{The Arrival of AI Native Software Engineering}

\vspace*{1mm}
\begin{quote}
\emph{“Life is like a box of chocolate.\\
You never know what you're going to get.”}\\
\hspace*{4mm}-- Forrest Gump
\end{quote}


By \emph{AI Native Software Engineering} (or \emph{AI-assisted}, \emph{AI-augmented} or \emph{AI-driven} software development), we refer to the use of AI tools and techniques to support, enhance, and automate various aspects of the software development lifecycle. This includes design, coding, debugging, testing, integration, deployment, orchestration and other typical software engineering tasks. AI models can provide real-time code suggestions, detect and fix errors, automate repetitive tasks, and even generate boilerplate code or entire modules and subsystems based on natural language descriptions. By analyzing patterns from existing codebases and understanding programming context, these tools can help developers improve productivity, reduce errors, and accelerate development time. In essence, AI-assisted development acts as an intelligent collaborator, making the software creation process more efficient and accessible. Such techniques can help less experienced developers act like true professionals, albeit there are also identified downsides \cite{ernst2022ai,waseem2024chatgpt-short}.

AI native software engineering and reuse relies on patterns learned from vast, diverse datasets of programs and documentation, including source code, documentation, tutorials, code snippets, and technical discussions (e.g., forums such as Stack Overflow or Reddit) \cite{ernst2022ai}. As such, software development based on AI generated code could also be referred to colloquially as {"\emph{second hand reuse}", since generated code is derived from numerous pre-existing sources that have been used as training material for the AI system.  
AI-assisted software reuse essentially constitutes a new form of generative reuse -- code is composed by AI, which has learned applicable development patterns from numerous earlier sources. While this approach is rapidly being adopted by developers, it has received surprisingly little attention from academic researchers until recently despite its potentially profound effect on software engineering practices.  


\begin{figure}[h]
    \centering
    \includegraphics[width=0.99\linewidth]{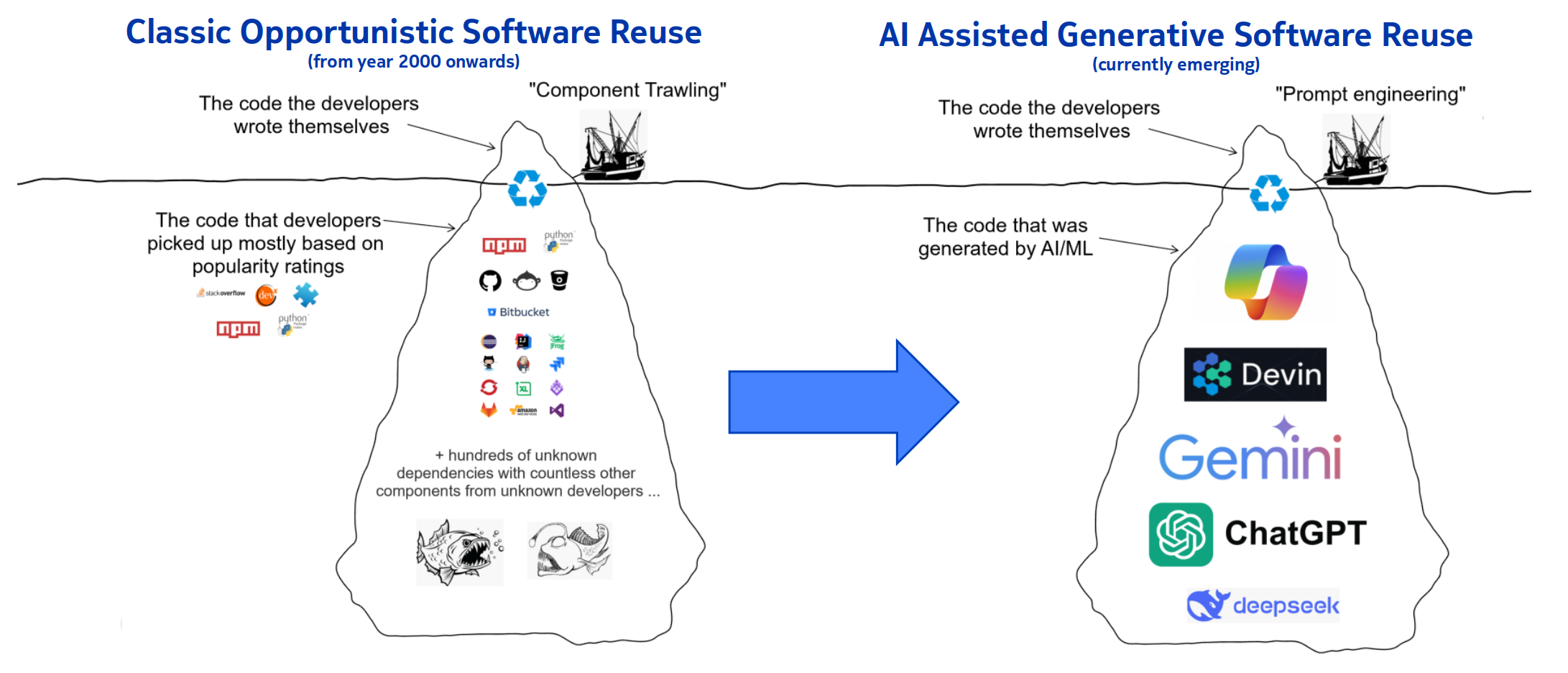}
    \caption{From Classic Opportunistic Software Reuse to AI Assisted Generative Software Reuse}
    \label{fig:recentevolution}
\end{figure}

Before diving further in AI native development practices, it should be noted that generative reuse is by no means an entirely new concept.  There are existing survey papers from over twenty-five years ago, e.g., by Biggerstaff \cite{biggerstaff1998generative}, and roots of this approach can be traced back to 1980s. In his paper, Biggerstaff analyzed success factors and challenges in software reuse, evaluated a number of then known generative reuse approaches, as well as provided measurements and metrics related to their effectiveness. There are also textbooks and various articles discussing generative programming long before the generative AI era \cite{czarnecki2000generativebook,czarnecki2004overview}. However, the concept of using AI for generating code and cobbling together entire solutions from various existing sources is still relatively new \cite{kapitsaki2024generative}.

As already stated above, AI native generative reuse is effectively a new form of opportunistic reuse in which the components "under the waterline" in the entire software "iceberg" are generated by artificial intelligence.  Unlike in classic opportunistic reuse, in which the underwater components have been selected from readily available component repositories by the developers themselves, in the AI driven approach the bulk of the system is generated automatically.  The approach used in generating those artifacts is sometimes colloquially referred to as \emph{"prompt engineering"} \cite{PromptEngineering2023,PromptEngineering2025} (Figure \ref{fig:recentevolution}) or \emph{"vibe coding"} \cite{sapkota2025vibecodingvsagentic}\footnote{The term "vibe coding" is attributed to Andrej Karpathy (\url{https://karpathy.ai/}).}.  While prompt engineering can be interpreted as a derogatory term, it is actually rather accurate since the approach that is used for generating the software components is based on rather loosely written textual prompts or queries that are entered and sent into the bowels of conversational AI assistants such as \emph{ChatGPT}, OpenAI \emph{Codex}, Microsoft \emph{Copilot}, Anysphere \emph{Cursor}, Google \emph{Gemini}, Perplexity or Replit \emph{Ghostwriter}.  While such an approach would have been completely unheard of only ten years ago, it is now becoming a reality that the majority of new software code can be generated in this fashion \cite{mikkonen2025fse}. In fact, some researchers go as far as argue that prompt engineering might ultimately replace the traditional requirement specification phase altogether.
The transition from classic opportunistic software reuse to AI assisted generative software reuse is depicted in Figure \ref{fig:recentevolution}.  

More broadly, prompt engineering and vibe coding refer to an emergent approach of software development in which the human programmer operates less as a direct implementer of code and more as a coordinator who communicates and collaborates with Large Language Models (LLMs) through iterative prompting. It has also been predicted that the iterative prompting phase might eventually be automated as well, leading us to \emph{agentic software development} or \emph{agentic coding} in which a significant part of system development is carried out in an entirely automated fashion \cite{sapkota2025vibecodingvsagentic}.  

At the time when this paper was written, Gartner placed \emph{Agentic AI} at the very peak of their 2025 hype cycle for generative artificial intelligence\footnote{\url{https://www.gartner.com/en/articles/hype-cycle-for-genai}}, and companies such as \emph{Lovable} (\url{https://lovable.dev/}) have received massive popularity for systems that claim to be able to create complex web-based applications and systems based on textual prompting alone; the long-term feasibility of such fully automated prompt-based software development still remains to be seen.




\section{Implications for Software Development}

The arrival of AI native software development has already resulted in a considerable mental model shift for software developers and for software engineering more broadly.  In this section we will summarize the claimed benefits, challenges and current limitations of this approach, as well as cover some existing studies on productivity gains.


\subsection{Claimed Benefits}

Commonly claimed benefits of AI native software development include areas such as productivity gains, code quality and error reduction, automated refactoring, maintenance and documentation. Below we provide an overview of the most common claims.

\begin{itemize}
\item Productivity Gains

\begin{itemize}
\item \emph{Faster task completion}. Several studies -- especially on tools such as GitHub Copilot -- report that developers complete coding tasks significantly faster when using AI tools. Estimates vary but figures such as 26\% (for nearly 5,000 developers across multiple companies) and 58\% (in controlled laboratory experiments) have been cited \cite{EffectsOfGenerativeAI2025}. Interestingly, as we will discuss a bit later in this paper, there are also contradictory studies that indicate significant productivity \emph{losses} when using AI tools.

\item \emph{Faster lead times especially for junior developers}. Several studies suggest that less experienced developers reap the most benefits from AI tools. For junior developers, AI can act as an always-available mentor that can help bridge skill gaps and accelerate learning of new development techniques and tasks.  In contrast, experienced developers seem to obtain productivity boosts for much more selective, targeted tasks, for instance, when starting to use a previously unfamiliar programming language, library or development framework.

\item \emph{Increased throughput}. According to studies, developers using AI tools tend to make more frequent code commits and successful builds, indicating a more streamlined development process.

\item \emph{Automation of repetitive tasks}. AI generally excels at automating mundane and repetitive coding tasks, freeing up developers to focus on more complex and creative aspects of software design and architecture. This includes generation of recommended programming patterns, boilerplate code, and even initial test case generation.

\end{itemize}

\item Code Quality and Error Reduction

\begin{itemize}
\item \emph{Error reduction}. AI tools can suggest corrections and improvements that will lead to cleaner, more efficient code with fewer errors. Some studies indicate a reduction in syntax errors and logical errors, although hallucination (discussed below) can wreak havoc in the generated outputs.

\item \emph{Improved test coverage}. AI-driven automated testing frameworks can generate test cases and execute them, enhancing testing accuracy and efficiency.

\item \emph{Assisted debugging}. AI can analyze errors and propose fixes automatically, accelerating bug resolution.
\end{itemize}

\item New Opportunities 

\begin{itemize}
\item \emph{Automated code generation}. Automated code generation is a prominent area, with LLMs generating code snippets, modules, subsystems, and even entire applications or end-to-end systems based on given textual input.

\item \emph{Predictive analytics}. AI can assist in analyzing project data to predict timelines, assess risks, and optimize resource allocation.

\item \emph{Refactoring and maintenance}. AI techniques can assist in suggesting optimal refactoring strategies and identifying anomalies in manual refactoring processes.

\item \emph{Automated documentation}. AI can automatically write technical descriptions, comments, examples and even tutorials from source code. This can be a significant productivity boon especially for smaller development teams.

\end{itemize}

\end{itemize}

\subsection{Challenges and Limitations}

The list below summarizes some of the key challenges and limitations associated with AI native software development.

\begin{itemize}
\item \emph{Accuracy and reliability}. While AI tools are powerful, they are by no means infallible. As a result of hallucination (discussed in more detail below), generated code may contain inaccuracies or inconsistencies that will require human verification and correction.

\item \emph{Context and nuance}. AI struggles with complex problem solving, or in understanding nuanced architectural requirements or ethical considerations that have traditionally required human judgment.

\item \emph{Ambiguity in requirement definition by prompt writing}. New inefficiencies can arise from the need for effective prompt writing and diligent verification of AI-generated solutions.  By nature, human languages such as English are much more ambiguous than programming languages.  Consequently, solution generation by prompt engineering is much more likely to result in ambiguous outputs that will require a lot of clarification and additional iterative prompting.  Furthermore, the results generated by such iterative prompting might not be repeatable, since LLM systems will commonly generate different outputs even when given exactly the same prompts as input.

\item \emph{Data privacy and intellectual property}. There are a lot of concerns related to the use of proprietary or copyrighted code for training AI models, or related to entering confidential information into AI tools when using those tools for development purposes. Furthermore, the use of AI tools may sometimes potentially invalidate patent claims, since the ownership rights of AI-generated code are not always clear.

\item \emph{Reduced problem-solving skills for human developers}. A lot of concerns exist that over-reliance on AI tools will have a significant negative long-term effects on the problem solving skills of human developers.  More junior developers might place their trust blindly on AI tools in a cargo cult like fashion without ever really understanding or learning what they are actually doing \cite{mikkonen2025fse}.  In general, the long-term impact of AI native development on human understanding and creativity has not yet been studied in detail.

\item \emph{Energy consumption}. While technical details related to the underlying computing machinery are usually hidden from the users of AI tools, LLMs typically require massive amounts of computing power. According to public sources, data centers in the US used somewhere around 200 terawatt-hours of electricity in 2024 -- roughly what it takes to power the entire country of Thailand for a year. AI-specific servers in these data centers are estimated to have used between 53 and 76 terawatt-hours of electricity. On the high end, this is enough to power more than 7.2 million US homes for a year. It is estimated that 80–90\% of computing power for AI is used for inference.  As the popularity of AI-based solutions increases, energy consumption demands are expected to grow dramatically.



\end{itemize}

\noindent \textbf{Notes on hallucination}. One of the biggest unresolved challenges in LLM-based AI development tools that needs to be highlighted is \emph{hallucination}, i.e., generation of incorrect or completely bogus results. In the broader scheme of things, LLM-based AI development tools and models are nothing more than glorified large-scale \emph{"stochastic typewriters"} or \emph{"stochastic parrots"}\footnote{\url{https://en.wikipedia.org/wiki/Stochastic_parrot}} \cite{BenderParrot2021}. Although some AI researchers dispute this characterization, LLMs are by and large limited by the data they are trained with, and they are simply stochastically repeating the contents of their input datasets. Because they are just making up outputs based on training data, LLMs do not really understand if they generate something incorrect or inappropriate.  In general, the models themselves do not -- and cannot -- decide what is true and what is false.  More broadly, hallucination is a built-in characteristic that arises from the very nature of the LLMs. It has been stated that especially with poor quality datasets, a learning machine might produce results that are "dangerously wrong" \cite{BenderParrot2021,LindholmBook2022}.

According to studies, hallucination rates with current AI development tools range from 1-3\% to nearly 80\% depending on the intended use case and domain.  One area in which current AI tools seem to behave especially erratically is in the area of scientific reference generation; when asked to provide references to scientific publications in a given technical area, current LLMs very commonly generate results that are either partially or sometimes completely bogus. At the first glance these generated references might look entirely convincing and correct, while specific details such as author or publication names could be completely contrived and incorrect.


In the area of code generation, LLM-based tools generally cope much better, but human developers must still be prepared to pinpoint and resolve occasional glitches that arise, e.g., from AI accidentally "grokking" and picking up parts of the generated solution from a development framework or a piece of code that is unrelated to the original request.


\subsection{Existing Studies on Productivity Gains}

There are significant variations among studies on the effectiveness of AI tools for software development.  Results vary from significant productivity increases to substantial productivity losses.  There are also studies that indicate no statistically significant task completion time improvements at all \cite{Vaithilingam2022}.


\textbf{Reports indicating productivity gains}. Cui et al. \cite{EffectsOfGenerativeAI2025} completed industry studies in 2022-2024 in which they analyzed three large-scale randomized controlled trials at Microsoft, Accenture and a third unnamed electronics manufacturing company.  In this study, Copilot coding assistant was made available to nearly five thousand software developers for a period of two to eight months in order to evaluate the effectiveness of the tool for their everyday work.  The results of the study indicated a 26\% increase in the weekly number of completed tasks for those using the tool, 13 1/2\% increase in the
number of code updates (commits) and over 38\% increase in the number of times code was compiled.  They also found Copilot availability to significantly raise task completion for more recent hires and those in more junior positions, thus reconfirming the results on the effectiveness of AI tools on less tenured employees in other technical areas \cite{Noy2023,Brynjolfsson2025}.

In an earlier study, conducted in a more laboratory-like setting, Peng et al. \cite{Peng2023} reported even more significant improvements.  In this trial, programmers were tasked to implement an HTTP server in JavaScript as quickly as possible. The trial group had access to GitHub Copilot and watched a brief video explaining how to use the tool. The control group did not have access to GitHub Copilot but was otherwise unconstrained, i.e., they were free to use internet search and Stack Overflow to complete the task.  In the results of the study, Peng et al. reported a 58\% decrease in time to complete a software engineering task in the lab, which would correspond to almost twice as many tasks done in a given amount of time.


Studies such as this will clearly have to be taken with a grain of salt, since the background of the developers who were chosen into the study groups would clearly have a significant impact on the results. For instance, any developer previously exposed to Node.js or other cloud backend development tools would have a dramatic advantage in a study such as this.  

A recent controlled study carried out by Alanazi et al. made a similar remark after carrying out 35 studies on the impact of AI tools on learning outcomes in computer programming courses \cite{Alanazi2025}. These studies demonstrated that students who utilized AI tools consistently outperformed those relying solely on traditional instructional methods, completing tasks more efficiently and producing higher-quality code.  However, while AI tools \emph{on average} had a positive effect on learning outcomes, there were significant variations indicating that AI effectiveness depends on various contextual factors, including course design, AI functionality, and student interaction \cite{Alanazi2025}.






\textbf{Reports indicating productivity losses}.
One recent randomized controlled trial published by Becker et al. in July 2025 found that AI tools surprisingly made experienced open-source developers take 
\emph{longer} to complete tasks, despite developers themselves expecting a significant speedup. Citing Becker et al. \cite{Becker2025}: 
\emph{
"When developers are allowed to use AI tools, they take 19\% longer to complete issues -- a significant slowdown that goes against developer beliefs and expert forecasts. This gap between perception and reality is striking: developers expected AI to speed them up by 24\%, and even after experiencing the slowdown, they still believed AI had sped them up by 20\%."} This highlights that while benchmarks show impressive AI capabilities, real-world usage on complex tasks with high quality standards might present different results.


The results vary considerably also based on which AI tools are used.  In a coding test performed by Zdnet in June 2025\footnote{\url{https://www.zdnet.com/article/the-best-ai-for-coding-in-2025-including-a-new-winner-and-what-not-to-use/}}, fourteen different AI development tools and LLMs were compared. Several of them generated results that were outright wrong, and only five tools (ChatGPT Plus, Perplexity Pro, Microsoft Copilot, Google Gemini Pro and Claude 4 Sonnet) ended up on the list of recommended tools.  While that study was by no means performed in a rigorous academic fashion, the results are still indicative of the great variations among the tools and the fact that some tools can produce highly misleading or incorrect results.

\textbf{Additional remarks}.  It should be noted that AI-assisted software development tools have evolved very rapidly in recent years. For instance, GitHub Copilot was developed by GitHub in partnership with OpenAI, and it was available for technical preview in June 2021 and publicly in June 2022.  Since then, the tool has evolved considerably as it became integrated comprehensively into Microsoft's product portfolio. Given rapid progress in the state-of-the-art, any studies carried out before year 2023 are likely rather badly out of date. 

\textbf{Tentative conclusions}. 
All in all, the results of studies on the productivity gains provided by AI-assisted software development tools are still contradictory.  Academic research so far indicates that AI tools can indeed be effective in many aspects of software development, primarily by automating routine tasks, improving code quality, and speeding up development cycles. However, the exact impact can vary, and challenges related to accuracy, human oversight, and the evolving demands in development skills still need to be addressed.  

The general conclusion from studies published so far seems to be that junior developers tend to benefit more from AI tools.  Nevertheless, because of the built-in tendency of LLM-based AI tools to hallucinate and generate occasional incorrect results and details, human experts are still required in various parts of the software development process to ensure the quality and correctless of the resulting system.  In short, \emph{AI tools lower barriers but cannot replace deep technical expertise.  More broadly, the future points towards collaborative intelligence where AI and human experts work together to leverage their respective strengths}.

\section{The Road Ahead -- Towards a Research Agenda}


\vspace*{1mm}
\begin{quote}
\emph{“If you don't know where you are going,\\
any road will get you there.”}\\
\hspace*{4mm}-- Lewis Carroll, \emph{Alice in Wonderland}
\end{quote}

At the time of this writing, it is still unclear what the long-term impact of AI native software reuse will be.  AI as a theme certainly has a cult-like following and "aura of omnipotence" around it at the moment -- people are putting a lot of faith on the ability of AI to magically accomplish almost anything. In their annual technology hype cycle in 2025, Gartner places "prompt engineering" and "AI augmented software engineering" very close to the top of the hype curve for generative AI, and "agentic AI" at the very peak\footnote{\url{https://www.gartner.com/en/articles/hype-cycle-for-genai}}.  Startup companies such as \emph{Lovable} (\url{https://lovable.dev/}) are currently capitalizing on this trend, and have recently received massive market valuations on the promise of prompt-based, fully agent-automated software system creation, although at the moment such systems tend to work only with specific target environments, languages and application frameworks -- such as TypeScript/React/Vite/Tailwind in case of the Lovable system.

In general, a lot of people in the software development community at large seem to have a fervent belief in the prompt engineering capabilities of generative AI.  Having witnessed the power of prompting capabilities in areas such as automated image and video generation -- including the use of Gaussian splatting techniques \cite{kerbl3Dgaussians} to generate vivid 3D scenes and even live videos out of conventional 2D photographs -- many people assume that in the area of software development it would ultimately be possible for AI to generate not only simple application skeletons but complete end-to-end software systems simply by providing detailed-enough textual paragraphs describing the requested system.  

In the ultimate scenario, traditional requirements definition phase would be skipped altogether based on the assumption that agentic AI could "fill in the missing gaps" automatically.  To us, \emph{such agentic prompt engineering approach would represent the ultimate form of cargo cult development}.  


Regardless of the ultimate outcomes, the arrival of AI native software engineering has already opened up numerous interesting research topics for academic researchers.  We have recently started collecting such topics systematically in order to form a research agenda for our research teams. 

Relevant questions in this area include:

\begin{itemize}
\item Can AI realistically generate complete, fully functional end-to-end systems?
\item What are the limits of prompt engineering in generating real life systems?
\item Can prompt engineering replace the requirement specification phase and other traditional phases in the software development lifecycle? 
\item Are human languages such as English, Finnish or Italian luculent/unambiguous enough for specifying requirements for AI generated systems, and/or ultimately replacing traditional programming languages altogether?
\item Is generative reuse an acceptable approach for generating code for production system use?
\item If the use of generative reuse becomes prevalent, can it be turned into a systematic method instead of the current \emph{ad hoc} ("let's try and see what AI can come up with") approach?
\item More broadly, is it possible to define systematic practices for generative AI-assisted reuse, or are the notions of a systematic method and generative AI fundamentally in conflict?
\item What is the equivalent of institutional reuse in the context of artificially generated code?
\item Can AI generate code that is maintainable and that adheres to specific coding standards and conventions?
\item How well can AI maintain and automatically modify the system when revised requirements/prompts are provided after taking the system in use?
\item How can requirement changes be tracked if the AI system generates different outputs even when given exactly the same prompts as input?
\item Can AI help to comprehend generated code, e.g., to articulate the role of specific lines of code in the broader context of the entire system (in addition to merely proposing code snippets to be used in specific locations in the code)?
\item How do we train human developers to evaluate the quality of artificially generated code?
\item How can developers evaluate quality of the generated artifacts if there are no natural subsystem boundaries or if the generated code does not follow any typical development patterns used by human developers?
\item How do we address, e.g., the copyright and security concerns associated with generative AI-assisted reuse?
\item Can AI fully automate maintenance, deployment, documentation, and other aspects related to reuse activities that are not directly associated with software system implementation?
\item Can AI be trained to take into account regulatory requirements that are mandatory in specific types of software systems and devices, e.g., such as ISO 13485 and IEC 62304 for the medical domain?  Can AI-generated code be certified at all for such use?
\end{itemize}



Regarding the first three questions in the list above, one of the biggest research themes to us at the moment is finding the realistic bounds and limits of prompt engineering.  Based on our experiences with AI-assisted software development tools so far, prompt engineering systems today could be characterized with an "80/20 rule": These systems can relatively easily propose code and generate skeleton applications for meeting roughly 80\% of the requirements.  However, finalizing the remaining $\sim$20\% of the system to specific requirements will require significant human oversight, tuning, reconciling and iteration in order to ensure that the resulting system is actually suitable for the intended domain and use.  This tuning phase can easily end up consuming $\sim$80\% of the development time.  While we have not yet performed any truly scientific empirical studies on actual percentages, this observation seems to hold true for those development activities that we have carried out so far.


\section{Discussion and Additional Comments}

Following up on the questions listed above, generative AI-assisted software reuse poses various additional challenges.  In particular, given that the innards of generative AI systems are effectively "sealed mystery boxes" to the users, developers using AI-generated code must be skilled enough to distinguish and evaluate the quality of the generated artifacts. While this may still be possible when the generated amounts of code are small, the situation can become unmanageable even for experienced software developers when entire end-to-end systems are being generated in an automated fashion. Merely blindly trusting whatever was automatically generated cannot be an acceptable approach if or when millions of lines of code are generated without human oversight or reviews.  Key concern here is that reuse based on assets that have not been proven to be fit for purpose or designed with reuse in mind can lead to serious quality and safety issues -- especially when those assets are used 
outside their original context. 
This is in sharp contrast with more classical forms of reuse that are assumed to improve software quality \cite{gao2003testing,trendowicz2003quality}.

Already back in 1998, Biggerstaff noted that an optimal strategy for generative reuse appears to favor systems in which there is a clear subsystem structure and in which those subsystems are generally organized in such a fashion that the size of an individual subsystem is in the KLOC range, i.e., size of individual subsystems and modules is only a few thousands of lines \cite{biggerstaff1998generative}.  AI-generated code might not follow such conventions; instead it could produce a monolithic system that satisfies the immediate requirements but which does not follow any traditional conventions that help improve system maintainability in the long term. This can introduce further mismatches with conventional approaches that are used in system testing, debugging, and deployment.

Furthermore, it should be noted that with AI-assisted development,
developers are ultimately reusing fragments of the training datasets. Therefore, developers should keep their eyes open for not only direct but "tangential" copyright and license term violations; while generated code might not be 100\% identical with known sources, it might still be close enough to trigger copyright and license conflicts. Besides quality and safety concerns, copyright issues are generally an interesting concern, since the ownership rights of AI-generated code are not always entirely clear.  Nowadays, there are tools such as \emph{Black Duck SCA} (see \url{https://www.blackduck.com/}) to detect and analyze such indirect copyright and license violations.

Code generation brings a new perspective to reuse since the code meant to be reused gets (automatically) rewritten over and over. This is similar to reuse by forking and duplicating the code as opposed to referring and linking to the same code meant to be reused. In the small, automated code completion has replaced the need to inefficiently copy and paste small code snippets, as such snippets can get automatically generated. The structure of the generated -- or reused, depending on the perspective -- code tends to be flat or undifferentiated, with many redundant code repetitions. This is typical of reuse by copy and paste which has empirically validated significant implications on the long-term maintainability of code \cite{narasimhan2015copy}.

Sometimes the generated code may also automatically refer to and dynamically import libraries that will be invoked and therefore reused by the generated code. These generated import statements may blindly assume that the library exists and that its interface is compatible with the generated code. If the reused library interface has evolved since the AI model was trained, the generated code will make use of deprecated features while ignoring the more recently added ones. In other cases, the name of the library referenced from the generated code might be hallucinated. The AI system might recommend reuse of code that is packaged in a library that does not exist anymore or at all (until someone goes and implements a new library with the generated name). Informally, problems in this area are referred to as \emph{slopsquatting}\footnote{\url{https://en.wikipedia.org/wiki/Slopsquatting}}.  Slopsquatting opens up a new types of supply chain attacks that do not require and depend on a takeover of an existing popular package or module but that rely on registering a new, potentially malicious packages under the hallucinated names.  The problems of generative reuse with deprecated libraries have been discussed in a number of papers, e.g., \cite{wang2025llmsmeetlibraryevolution}.

More broadly, security concerns related to the combination of LLMs and agentic AI capabilities cannot be ignored.  These concerns arise from the ability of agentic AI systems to leverage external systems by invoking, e.g., a Python or JavaScript code (and indirectly native code) to perform different types of tasks on behalf of the user in the form of remote function calls. Furthermore, many LLMs are nowadays multimodal and capable of responding not only to English text but also to images, audio and various other types of input (in numerous human languages, programming languages and various other notations and file formats).  Some researchers have pointed out that these multimodal prompt-based remote code execution capabilities can be utilized to perform malicious tasks as well as reveal personal details about unrelated to users and software systems in an especially cryptic, undetectable fashion \cite{fu2024impromptertrickingllmagents}.  These capabilities can also be potentially be used to jailbreak software code that should not be accessible externally \cite{liu2024jailbreakingchatgptpromptengineering}.  While the broader security concerns related to agentic AI are beyond the domain of software reuse, this area definitely deserves further research as well.

\section{Conclusions}

Cumulatively, the three authors of this paper have over a hundred years of expertise in the software industry both from academic and industrial perspectives.  Over time, we have witnessed and studied the implications of several disruptions that have occurred the software industry and in the software engineering field more broadly.  
We have also systematically studied software reuse and its evolution over the decades, as well are built significant software systems ourselves both as producers and consumers of reusable software.

Historically, more significant paradigm shifts in the computing industry have occurred approximately every 10-15 years.  
Out of the disruptions that we have observed, the current ongoing transition towards AI native software engineering might very well turn out to be the most dramatic and impactful paradigm shifts ever.  We are rapidly moving from "organic software", i.e., organically produced, humanly written software to machine-generated software.  

At the same time, AI-generated and AI-driven software is increasingly given access rights to control physical real-world infrastructure and systems such as factories, office equipment, home appliances, automobiles, and smart city infrastructure.  This new approach has emerged so quickly that the industry and academic researchers are yet to absorb and fully catch up with the impact.  In those software development projects that we have been involved in the past four years, we have seen the use of AI tools increase dramatically, and it is rather clear by now that the general approach is here to stay. 

In this paper, we have taken a look at the emerging era of AI native software engineering especially from the viewpoint of software reuse.  We started the paper with a condensed history of software reuse in the past sixty years.  We commented on how software development progressed from fully in-house written software from the late 1960s, 1970s and early 1980s to popular commercially available component frameworks in the late 1980s and early 1990s, prior to the era of large-scale opportunistic software reuse enabled by the World Wide Web in the turn of the century.  In the past two decades, software developers have rarely written any code fully from scratch given the massive amount of readily available software components in popular software repositories such as GitHub, Node Package Manager and Python Package Index.

In the remaining parts of the paper, we took a look at the AI-assisted software development approach that is now rapidly surpassing and replacing the previously dominant opportunistic software reuse model.  We took at look at the implications of this new approach, including its benefits, current limitations and existing studies on productivity gains. We then put together a tentative research agenda summarizing those research questions that seem most relevant to us, followed by some discussion and general concerns related to prompt engineering and AI native software development more broadly. 

While the final bounds of AI native software are yet to be discovered, current research points towards collaborative intelligence where AI and human experts work together to leverage their respective strengths.  It remains to be seen whether the boundaries can be pushed even further, with machine-generated software eventually replacing human developers altogether.




\bibliographystyle{plain}

\bibliography{bibliography}

\end{document}